\documentclass{jpp}
\usepackage{graphicx}
\usepackage{epstopdf, epsfig}
\usepackage{hyperref}
\usepackage{amsmath,amssymb}
\usepackage[pdftex,dvipsnames]{xcolor}  

\def\be{\begin{equation}}
\def\ee{\end{equation}}

\def\der#1#2{{\partial #1\over \partial #2}}

\shorttitle{A formal equivalence between plasma and rotating shallow water waves}
\shortauthor{E. Heifetz, L. R. M. Maas, J. Mak and I. Pomerantz}

\title{On a formal equivalence between electro-magnetic waves in cold plasma and
shallow water inertio-gravity waves}

\author{E. Heifetz\aff{1},
  Leo R. M. Maas\aff{2},
  J. Mak\aff{3}
  \corresp{\email{julian.c.l.mak@googlemail.com}} \and I. Pomerantz\aff{4} }

\affiliation{\aff{1}Porter school of the Environment and Earth Sciences, Tel Aviv University,
 69978, Israel
\aff{2}Institute for Marine and Atmospheric research Utrecht, University of Utrecht, 3584 CC Utrecht, NL
\aff{3}Dept. of Ocean Science and Center for Ocean Research in Hong Kong and Macau, Hong Kong University of Science and Technology, Clearwater Bay, Hong Kong SAR
\aff{4}School of Physics and Astronomy, Tel Aviv University, 69978, Israel
}

\begin{document}

\maketitle

\begin{abstract}
The fundamental dispersion relation of transverse electro-magnetic waves in a cold collisionless plasma is formally equivalent to the two dimensional dispersion relation of inertio-gravity waves in a rotating shallow water system, where the Coriolis frequency can be identified with the plasma frequency, and the shallow water gravity wave phase speed plays the role of the speed of light. Here we examine this formal equivalence in the governing linearised equations, and compare between the propagation wave mechanisms in these seemingly unrelated physical systems.

\end{abstract}


\section{Formal equivalence in the wave dispersion relations}

Consider a cold collisionless plasma whose ions are assumed to be at rest. The electron mean dynamics is then governed by the charge continuity and momentum equations \citep{bittencourt2013fundamentals}:

\be
\der{n}{t} = -\nabla\cdot(n {\bf u})\, ,
\label{charge_cont}
\ee
\be
\der{{\bf u}}{t} = -({\bf u}\cdot\nabla){\bf u} -{e\over m}({\bf E}+ {\bf u}\times{\bf B})\, ,
\label{momentum}
\ee
and Maxwell's equations:
\be
\der{{\bf B}}{t} = -\nabla\times {\bf E}\, ,
\label{Farady}
\ee
\be
\der{{\bf E}}{t} = c^2\,\nabla\times {\bf B} + {e\, n \over \epsilon_0}{\bf u}\, ,
\label{Ampere}
\ee
\be
\nabla\cdot {\bf E} = {\rho
\over \epsilon_0} = {q_i n_i -e \, n\over \epsilon_0}\, ,
\label{Gauss}
\ee
\be
\nabla\cdot {\bf B} = 0\, ,
\label{GaussMagnetic}
\ee
where $t$ denotes the time, $\nabla$ is the 3D nabla operator, $n$ is the electron number density, ${\bf u}$ is the mean electron velocity, $e$ and $m$ are the electron charge and mass,  ${\bf E}$ and ${\bf B}$ are the electric and magnetic fields, $c$ is the speed of light, $\epsilon_0$  the vacuum permittivity, $\rho$ the total electric charge density, and $q_i$ and $n_i$ are the ion charge and number density. These equations admit an equilibrium state with ${\bf E} = {\bf B} = {\bf u} = \boldsymbol{0}$ and $\rho = 0$, with an equilibrium electron number density $n_0 = q_i n_i/e$ (indicated by subscript zero) that is uniform in space and time. Denoting perturbations quantities by primes, the linearised equations about the rest equilibrium reads:
\be
\der{n'}{t} = -n_0\nabla\cdot{\bf u}'\, ,
\label{charge_contLin}
\ee
\be
\der{{\bf u}'}{t} = -{e\over m}{\bf E}'\, ,
\label{momentumLin}
\ee
\be
\der{{\bf B}'}{t} = -\nabla\times {\bf E}'\, ,
\label{FaradyLin}
\ee
\be
\der{{\bf E}'}{t} = c^2\,\nabla\times {\bf B}' + {e\, n_0 \over \epsilon_0}{\bf u}'\, ,
\label{AmpereLin}
\ee
\be
\nabla\cdot {\bf E}' = -{e \over \epsilon_0} n'\, ,
\label{GaussLin}
\ee
\be
\nabla\cdot {\bf B}' = 0\, .
\label{GaussMagneticLin}
\ee

This linear set of equations have some redundancies (the divergence of \eqref{AmpereLin}, with the use of  \eqref{GaussLin}, implies \eqref{charge_contLin}, and \eqref{FaradyLin} is consistent with \eqref{GaussMagneticLin}), and plane wave solutions of the form of $f'({\bf x},t) = {\hat f}e^{i({\bf k}\cdot{\bf x} - \omega t)}$ (where $f'$ represents any perturbed variable at position ${\bf x}$, and time $t$, with wavevector ${\bf k}$ and angular frequency $\omega$) can be described solely from the equation subset \eqref{momentumLin}-\eqref{AmpereLin}. The linear wave solutions are of two types. One is when the electric field is non-rotational ($\nabla \times {\bf E}'=0$, implying that the wavevector is aligned with the perturbation electric field, as ${\bf k}\times {\bf E}'=0$), so the perturbation magnetic field is zero, and the dispersion relation is $\omega = \pm \omega_p$, where $\omega_p \equiv \sqrt{n_0 e^2 / m\, \epsilon_0}$ is the plasma frequency for the present setup. 
The second is the transverse solution when the electric field is solenoidal (i.e. non-divergent, $\nabla \cdot {\bf E}'=0$, 
implying that the wavevector is perpendicular to the perturbation electric field, as ${\bf k}\cdot {\bf E}'=0$), so the perturbation electron number density $n'$ vanishes, with dispersion relation $\omega^2 = \omega_p^2 + (kc)^2$, where $k = |{\bf k}|$. 

\begin{figure}
    \centering
    \includegraphics[width=1\textwidth]{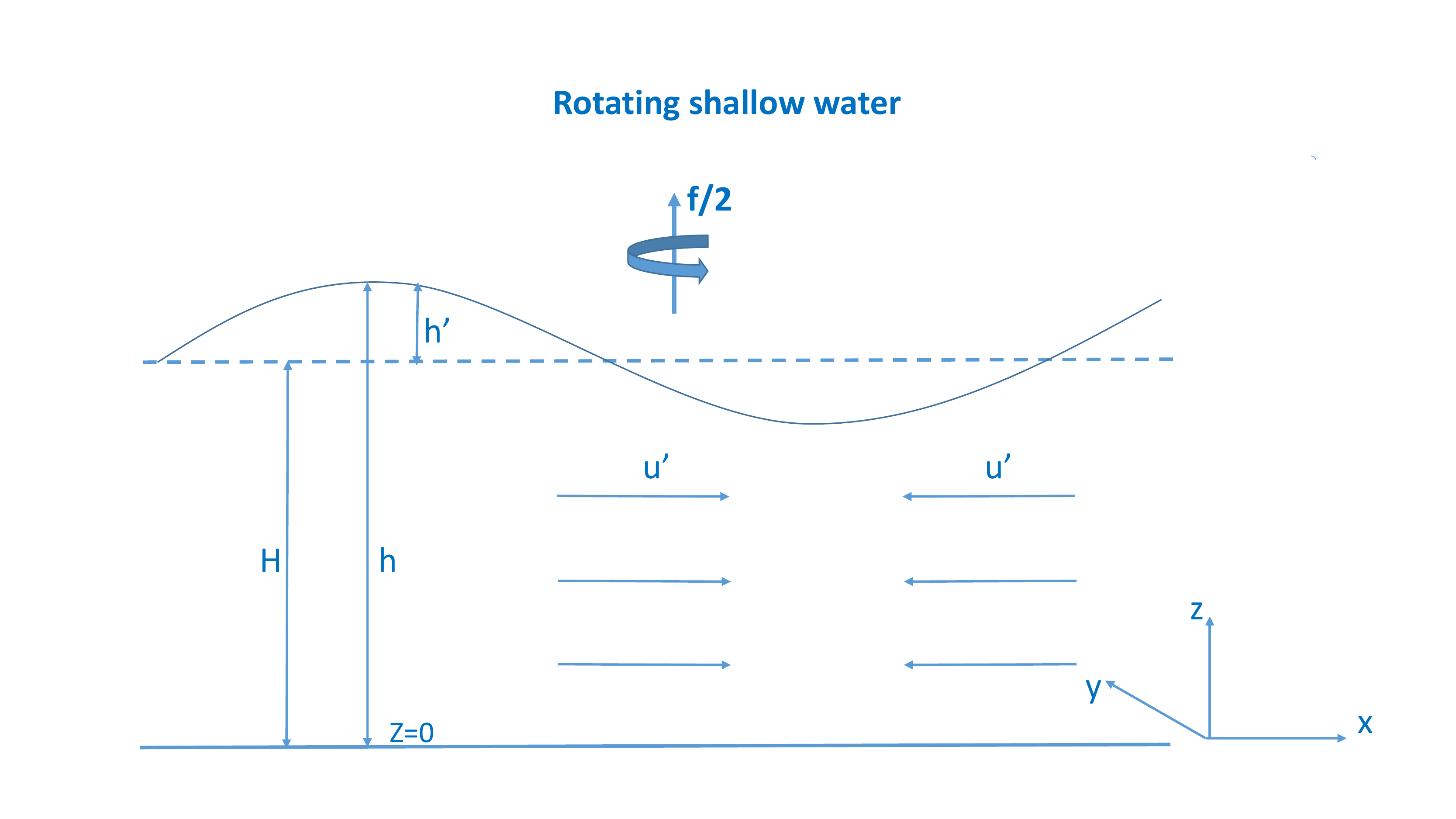}
    \caption{Schematic diagram of the rotating shallow water system. The system rotates counterclockwise in the horizontal direction with an angular velocity of $f/2$, pointing upwards, where the gravity field ${\bf g}$ is pointing downwards. The total layer depth is $h(x,y,t) = H +h'(x,y,t)$. The horizontal velocity field is independent of $z$.}
    \label{fig:1}
\end{figure}

Consider in turn the uniformly rotating shallow water (RSW) system (Fig.\,~\ref{fig:1}) of an incompressible (neutrally charged) fluid \citep{vallis}, described by the continuity and momentum equations:  
\be
\der{h}{t} = -\nabla_H\cdot(h {\bf v})\, ,
\label{SW_cont}
\ee
\be
\der{{\bf v}}{t} = -({\bf v}\cdot\nabla_H){\bf v} -g\nabla_H h +{\bf v}\times{\bf f}\, .
\label{SW_momentum}
\ee
Here $h(x,y,t)$ denotes the layer's height, ${\bf v}(x,y,t) = (u,v)$ is the depth-independent horizontal velocity vector within the fluid layer, $\nabla_H$ is the horizontal gradient operator and ${\bf f} = f {\hat{\bf z}}$ is the Coriolis frequency pointing upwards (${\hat{\bf z}}$ is the vertical unit vector), where we take $f = \textnormal{constant}$. The latter is equal to twice the rotation frequency of the system ($f$ is positive for counter-clockwise rotation). The term $-g\nabla_H h$ represents the horizontal pressure gradient force (as the fluid is assumed to be in hydrostatic balance), thus horizontal pressure differences within the layer results solely from differences in the layer's height. The term ${\bf v}\times{\bf f}$ represents the Coriolis force, and it acts to the right (left) of the flow's motion for positive (negative) values of $f$.

Assuming no topography, equations \eqref{SW_cont}-\eqref{SW_momentum} admit a rest state solution of mean constant height $h=H$. Again, denoting the deviation quantities by primes, the linearised RSW equations reads:
\be
\der{h'}{t} = -H\nabla_H\cdot{\bf v}'\, ,
\label{SW_cont_Lin}
\ee
\be
\der{{\bf v}'}{t} =  -g\nabla_H h' +{\bf v}'\times{\bf f}\, .
\label{SW_momentum_Lin}
\ee
The linearised RSW system supports horizontal wave solutions, described by the geostrophic mode with dispersion relation $\omega = 0$, as well as inertio-gravity (Poincare) waves with dispersion relation $\omega^2 = f^2 + (k_H c_s)^2$, where $k_H$ is the magnitude of the horizontal wavevector, and $c_s=\sqrt{gH}$ is the shallow water surface gravity wave phase speed. Introducing the Rossby radius of deformation $L_d \equiv c_s/f$, the dispersion relation for the inertio-gravity waves can be rewritten as $\omega^2 = f^2[1+ (k_H L_d)^2]$. Thus, in the short wave limit $k_H L_d \gg 1$, we obtain gravity waves with $\omega \rightarrow \pm k\, c_s$, whereas in the long wave limit $k_H L_d \ll 1$, we obtain inertial oscillations with $\omega \rightarrow \pm f$, a low-frequency cut-off equivalent to $\omega_p$.


\section{Formal equivalence in the linearised equations}

The similarity in the dispersion relations, between electro-magentic waves in a cold plasma $\omega^2 = \omega_p^2 + (kc)^2$ and inertio-gravity waves in the rotating shallow water system $\omega^2 = f^2 + (k_H c_s)^2$, is intriguing, given that the system possesses different physics, and that the linearised RSW system for the three scalar perturbation variables $(h, u, v)$ (omitting the primes hereafter for the perturbation variables)
\be
\der{h}{t} = -H\left ( \der{u}{x} + \der{v}{y} \right)\, ,
\label{SW_cont_uv}
\ee
\be
\der{u}{t} =  -g\der{h}{x} +f v\, ,
\label{SW_momentum_u}
\ee
\be
\der{v}{t} =  -g\der{h}{y} -f u\, ,
\label{SW_momentum_v}
\ee
seem, at first sight, to have little in common with the three linearised vector equations \eqref{momentumLin}-\eqref{AmpereLin} for the three vector fields $({\bf u},{\bf B}, {\bf E})$. However, one can in fact obtain a formal equivalence between the two linearised systems as follows. Starting with the linearised RSW equations, writing it in terms of the divergence $\delta$, vertical component of the vorticity $\zeta$ (normalized by $f$) and the height perturbation (normalized by $H$) $B$, given by
\be
{{\delta}}\equiv \nabla_H\cdot{\bf v} = \left (\der{u}{x} + \der{v}{y}\right ), 
\hspace{0.25cm}
{{\zeta}}\equiv {1\over f}(\nabla_H\times{\bf v})\cdot{\hat{\bf z}} = {1\over f}\left (\der{v}{x} - \der{u}{y}\right ),
\hspace{0.25cm}
B \equiv  {h\over H},
\label{scalars_definition}
\ee
the linearised RSW equations \eqref{SW_cont_uv}-\eqref{SW_momentum_v} are transformed into
\be
\der{{\zeta}}{t} = -{{\delta}}\, ,
\label{zetaSW}
\ee
\be
\der{{B}}{t} =  -{{\delta}}\, ,
\label{BSW}
\ee
\be
\der{{\delta}}{t} = -c_s^2\,\nabla_H^2 {B} + f^2\, {{\zeta}}\, .
\label{deltaSW}
\ee

On the other hand, if we define in the plasma system
\be
{\pmb {\delta}}\equiv \nabla\times {\bf E},
\hspace{0.5cm}
{\pmb {\zeta}}\equiv {m\over e}\nabla\times {\bf u},
\label{vectors_definition}
\ee
then the system \eqref{momentumLin}-\eqref{AmpereLin} reads
\be
\der{\pmb {\zeta}}{t} = -{\pmb {\delta}}\, ,
\label{zetaplasma}
\ee
\be
\der{{\bf B}}{t} =  -{\pmb {\delta}}\, ,
\label{Bplasma}
\ee
\be
\der{\pmb {\delta}}{t} = -c^2\,\nabla^2 {\bf B} + \omega_p^2\, {\pmb {\zeta}}\, .
\label{deltaplasma}
\ee
Hence, within the linear framework, the RSW system formally resembles the plasma one with no background magnetic field (where $c_s \leftrightarrow c$, $f  \leftrightarrow \omega_p$), with the limitations that the former is of scalar variables and the latter is of vector ones. Furthermore, the gradient operator in the former is horizontal, whereas in the latter is fully three-dimensional. 


\section{Wave propagation mechanism in the formally equivalent systems}

\begin{figure}
    \centering
    \includegraphics[width=1\textwidth]{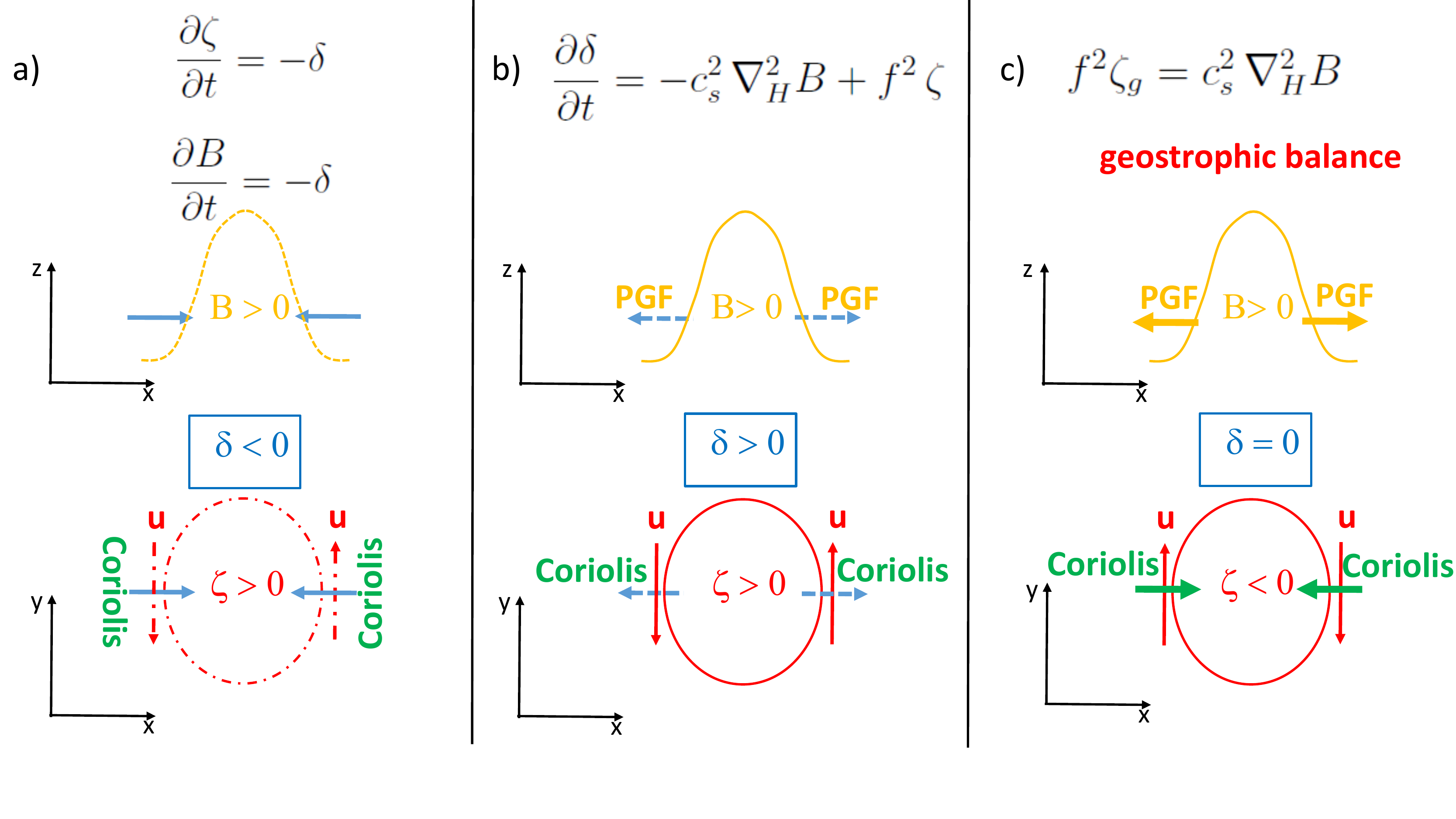}
    \caption{Schematic illustrating linearised dynamics in the RSW system. Solid arrows and curves indicate a current state where dashed lines and curves represent the system's response. a) Initial response to perturbations, b) feedback and c) steady state. Top panels recall governing equations; middle and bottom panels show vertical and horizontal cross-sections.  
    }
    \label{fig:2}
\end{figure}

\begin{figure}
    \centering
    \includegraphics[width=1\textwidth]{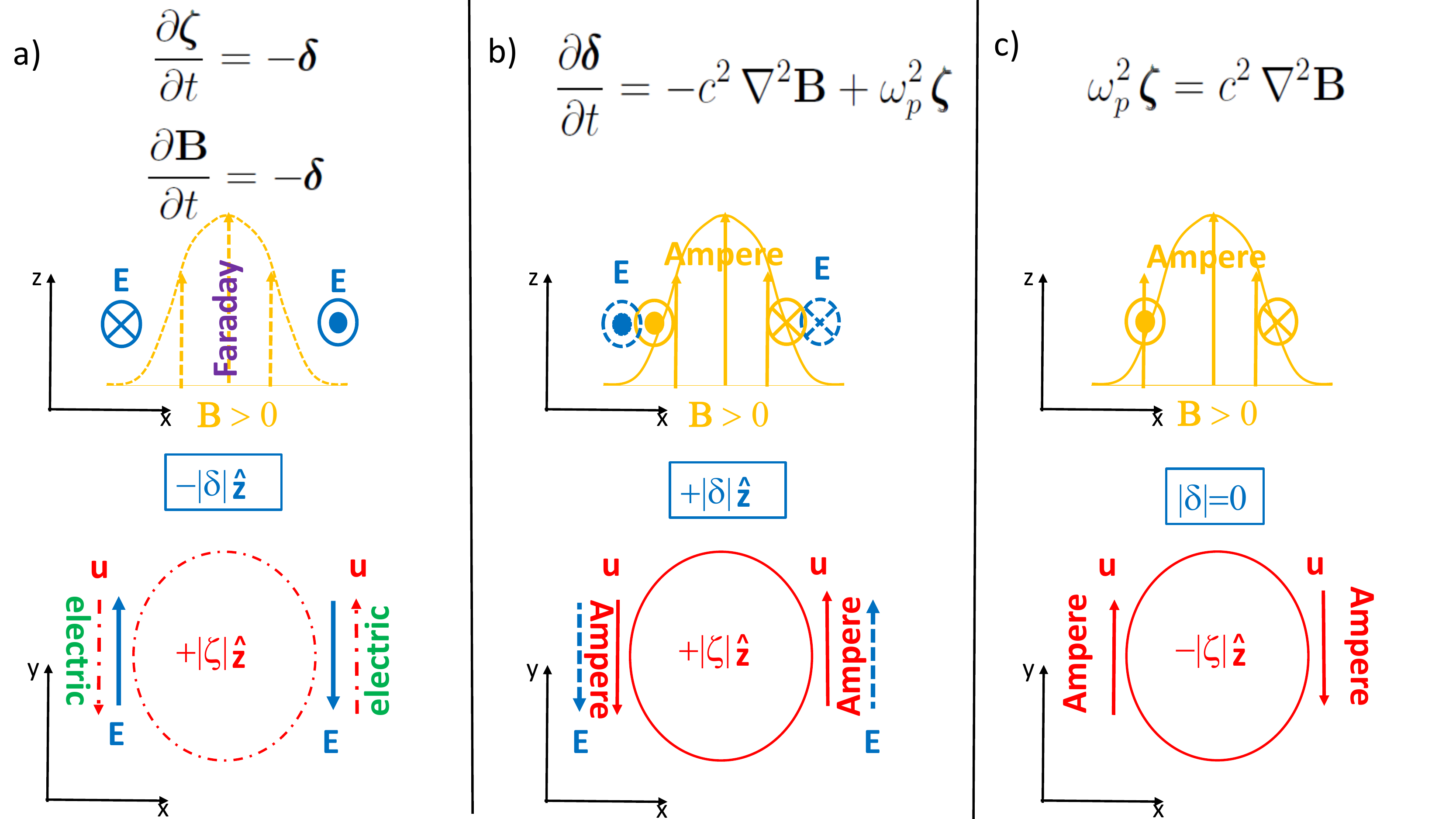}
    \caption{As Fig.\,~\ref{fig:2}, for the plasma system.}
    \label{fig:3}
\end{figure}

We wish to mechanistically understand why the two systems share a similar mathematical formalism given there is no \emph{a priori} expectation for them to do so. This is schematically illustrated in Figs.~\ref{fig:2} and \ref{fig:3}. Consider first the linearised dynamics of the RSW system where equations \eqref{zetaSW}-\eqref{BSW}, repeated at the top of Fig.~\ref{fig:2}a, govern  the response to an initial perturbation consisting in convergence ($\delta<0$, solid blue arrows). Assuming $f$ is positive,   the Coriolis force (green thick arrows) acts to the right of the inward motion, and as \eqref{zetaSW} tells convergence leads to the generation of positive (counterclockwise) vorticity $\zeta>0$ (red dashed circle in bottom panel). As the shallow water system is incompressible, according to \eqref{BSW} horizontal convergence simultaneously lifts up the layer's height, yielding a positive height $B$ anomaly (dashed orange curve in middle panel). 

Consider in turn  the linearised dynamics in the plasma system where \eqref{zetaplasma}-\eqref{Bplasma}, repeated at the top of Fig.\,~\ref{fig:3}$a$, govern the response to an initial perturbation in the electric field. The latter is assumed to be horizontal and circulating clockwise (solid blue arrows  and  in and out vector signs). This implies the curl of the electric field, 
${\pmb \delta} = \delta{\hat {\bf z}}$, with $\delta<0$, is pointing downward. The electric field apply an electric force (green thick arrows) acting on the electrons on the opposite direction \eqref{momentumLin}. According to  \eqref{zetaplasma}  this generates counterclockwise flow circulation ${\pmb \zeta} = \zeta{\hat {\bf z}}$, with $\zeta>0$, is pointing upward (red dashed circle in bottom panel). Simultaneously, according to  Faraday's law in \eqref{FaradyLin} and \eqref{Bplasma}, the clockwise circulation of the electric field yields an upward pointing magnetic field (${\pmb B} = B{\hat {\bf z}}$, with $B>0$, orange line in middle panel).

Fig.\,~\ref{fig:2}$b$ shows that in the RSW system the new positive height and vorticity anomalies prompt a feed-back. As the positive height field anomaly yields a hydrostatic high pressure anomaly, both the pressure gradient force (PGF, magenta thick arrows) and the Coriolis force, acting  to the right of the rotational flow, lead to an outward divergent flow, against the initial convergence, generating divergence ($\delta>0$). 

In turn, Fig.\,~\ref{fig:3}$b$ shows that in the plasma system the new positive magnetic and vorticity anomalies similarly create a feed-back. They act together to generate counterclockwise circulation of the electric field (${\pmb \delta} = \delta{\hat {\bf z}}$ with $\delta > 0$), in agreement with \eqref{deltaplasma}. The Ampere--Maxwell equation \eqref{AmpereLin} indicates that both the curl of the magnetic field and the electron motion generate an electric field. The radial shear of the vertical magnetic field anomaly (indicated by the solid orange arrows in the vertical cross-section) yields a curl pointing counterclockwise in the azimuthal direction (represented by the magenta in and out vector signs), which in turn generates a counterclockwise electric field (blue in and out vector signs).    
The counterclockwise electron flow (red solid arrows) yields as well an electric field in the direction of their flow. Hence the vorticity ${\pmb \zeta}$ associated with the electron flow generates curl of the electric field ${\pmb \delta}$.

By combining the equations in the top panels of Figs.\,~\ref{fig:2}a and 2b, respectively \ref{fig:3}a and 3b, the initial perturbations and their feed-backs guarantee the presence of an oscillatory response in both systems. These represent waves propagating around a state in which $\zeta, B$ and $\delta$ and their vector equivalents vanish. At the end of this section we will interpret this state, as it turns out to be nontrivial. 

However, as Figs.\,~\ref{fig:2}$c$ and \ref{fig:3}$c$ show, these systems also support a nonzero equilibrium state (indicated by subscripts g). In the RSW system this is the celebrated geostrophic balance  between pressure gradient and  Coriolis forces (top panel), ${\bf v}_g = (g/f){\hat {\bf z}}\times \nabla_H h$. This is obtained in the absence of any radial currents, when a high pressure anomaly $B>0$ (orange arrows and in and out vector signs) is located at the center of a clockwise circulation anomaly ($\zeta < 0$). 
Consequently the geostrophic flow is non-divergent ($\delta_e=0$). 

As Fig.\,~\ref{fig:3}$c$ shows, the plasma system has a nontrivial  equilibrium too. This happens in the absence of an electric field,  where the Ampere--Maxwell equation \eqref{AmpereLin} satisfies ${\bf u}_g = (\epsilon_0 c^2/e n_0)\nabla\times {\bf B}$. Then ${\pmb {\delta}}_e = 0$ and the two terms in the RHS of  \eqref{deltaplasma} balance each other as $\omega_p^2\, {\pmb {\zeta}}_e = c^2\,\nabla^2 {\bf B}$. This can be achieved when electrons circulate in a horizontal clockwise direction and balance a magnetic field anomaly that points upward.

\begin{figure}
    \centering
    \includegraphics[width=1\textwidth]{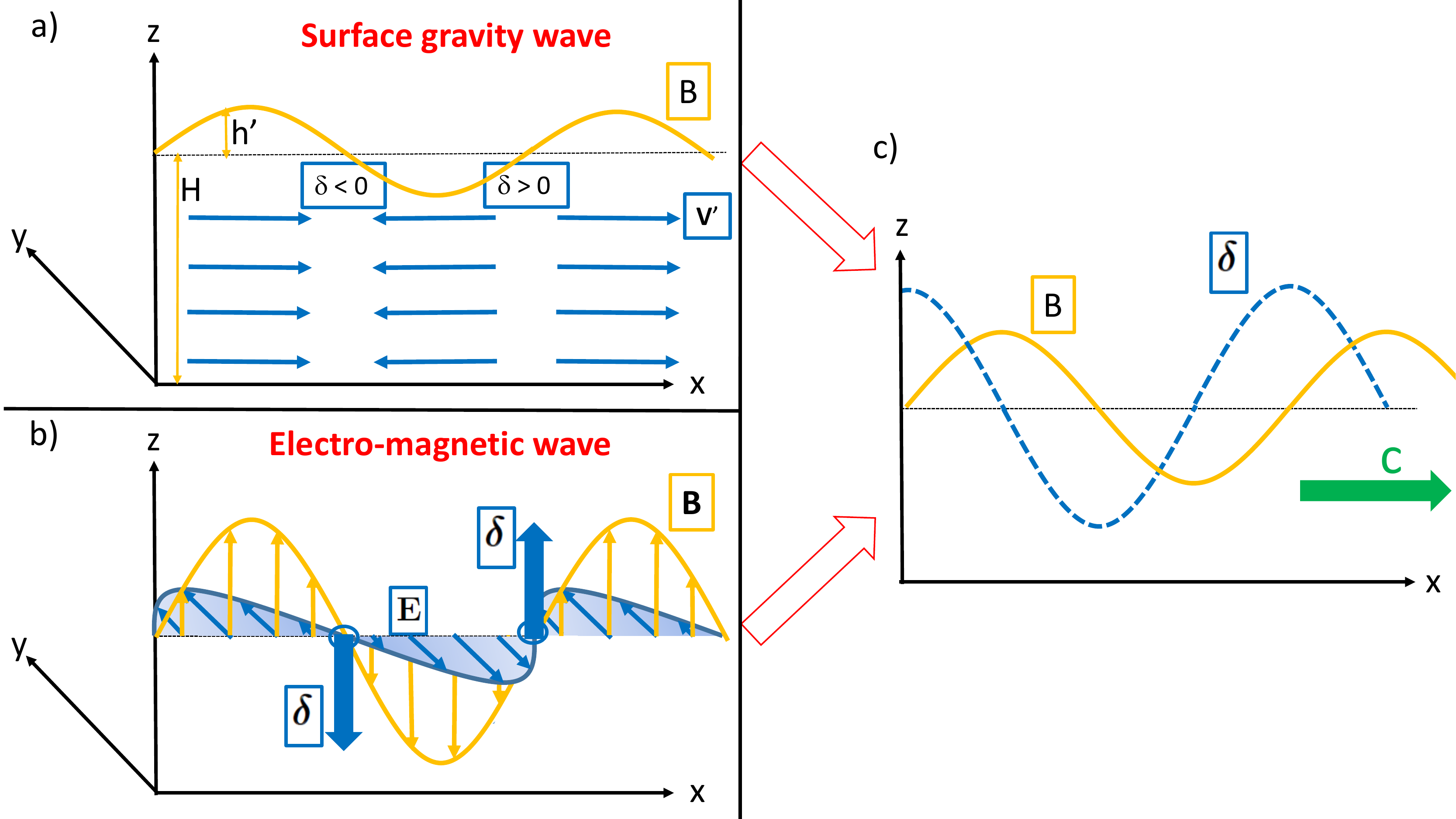}
    \caption{A comparison between the  propagation mechanisms of surface gravity waves in  non-rotating shallow water  and of electro-magnetic waves in vacuum. 
    ($a$) Horizontal velocity (blue arrows),  divergence $\delta$ and vertical displacement field (orange line)  in surface gravity wave. 
    ($b$) Electric (thin blue arrows) and  magnetic (orange arrows) fields and curl of electric field ${\pmb \delta}=\delta {\pmb e_z}$ (thick blue arrow).
    ($c$) The propagation mechanism of both types of waves expressed in terms of the   structure of scalar $B$ and $\delta$ fields, where ${\pmb c}$ indicates the phase propagation velocity.  
    }
    \label{fig:4}
\end{figure}

Consider now the short wave limit solutions of the RSW system with length-scales much shorter than the Rossby deformation radius. This is an exact limit for the case when the shallow water system is non-rotating ($f=0$) and the flow itself is irrotational. For this case, system \eqref{zetaSW}-\eqref{deltaSW} reduces to the solution of the non-dispersive surface gravity waves: 
\be
\der{{B}}{t} =  -{{\delta}},
\hspace{0.25cm}
\der{{\delta}}{t} = -c_s^2\,\nabla_H^2 {B}
\hspace{0.5cm} \Longrightarrow \hspace{0.5cm}
\omega^2 = (k_H c_s)^2\,  .
\label{GW}
\ee
For surface gravity waves, the propagation mechanism  can be understood, using Fig.\,~\ref{fig:2} and Fig.\,~\ref{fig:4}$a$, in terms of the interplay between horizontal divergence/convergence and the undulation of surface elevation. Convergence (divergence) generates ridges (crests) which in turn, by applying a pressure gradient force, generates divergence (convergence). 
The horizontal velocity and the vertical displacement fields are in phase. Consequently, the resulting pressure gradient force translates the velocity field to the right in concert with the height field that is translated by the horizontal convergence/divergence field. Hence, for rightward propagation the divergence field lags the height field by a quarter of a wavelength (for leftward propagation the height and the velocity fields are anti-phased).

It is interesting that this non-rotating  limit of  (irrotational) surface gravity waves corresponds to the limit of the plasma system \eqref{zetaplasma}-\eqref{deltaplasma} in a vacuum, (i.e., in the absence of electron flows, thus ${\pmb {\zeta}} = 0$), admitting the familiar electro-magnetic wave solution:
\be
\der{\bf {B}}{t} =  -{\pmb {\delta}},
\hspace{0.25cm}
\der{\pmb{\delta}}{t} = -c^2\,\nabla^2 {\bf B}
\hspace{0.5cm} \Longrightarrow \hspace{0.5cm}
\omega^2 = (k c)^2\,  .
\label{EM}
\ee
The electric and  magnetic fields are in phase, where the former is pointing in the $y$ direction and the latter in the $z$ direction, see Fig.\,~\ref{fig:4}b. The curl of the electric field is also pointing in the $z$ direction and its negative values translates the magnetic field rightward. In turn, the curl of the magnetic field, pointing in the $y$ direction, translates the electric field rightward as well (for leftward propagation the electric  and the magnetic  fields are anti-phased).

As illustrated in Fig.\,~\ref{fig:4}$c$, for the setups shown in Figs.\,~\ref{fig:4}$a$ and $4b$, despite the difference in physics both mechanisms converge into the same description in which the $\delta$ and the $B$ fields are wavy scalar fields in the $x$-$z$ plane, where the $\delta$ lags $B$ by a quarter of a wavelength and their amplitude ratio $|\delta|/|B| = |\omega|$. As $B$ generates $\delta$ and $-\delta$ generates $B$, the two fields propagate in concert in the positive $x$ direction. Furthermore, as for the speed of light in vacuum, the shallow water surface gravity waves are non-dispersive   and constitutes the largest possible attainable value of gravity wave phase and group speeds (for a given mean layer depth $H$).
This equivalence, between non-rotating shallow water surface gravity wave propagation in a {\em continuous medium}, and electro-magnetic wave propagation in {\em vacuum}, is intriguing.

\begin{figure}
    \centering
    \includegraphics[width=1\textwidth]{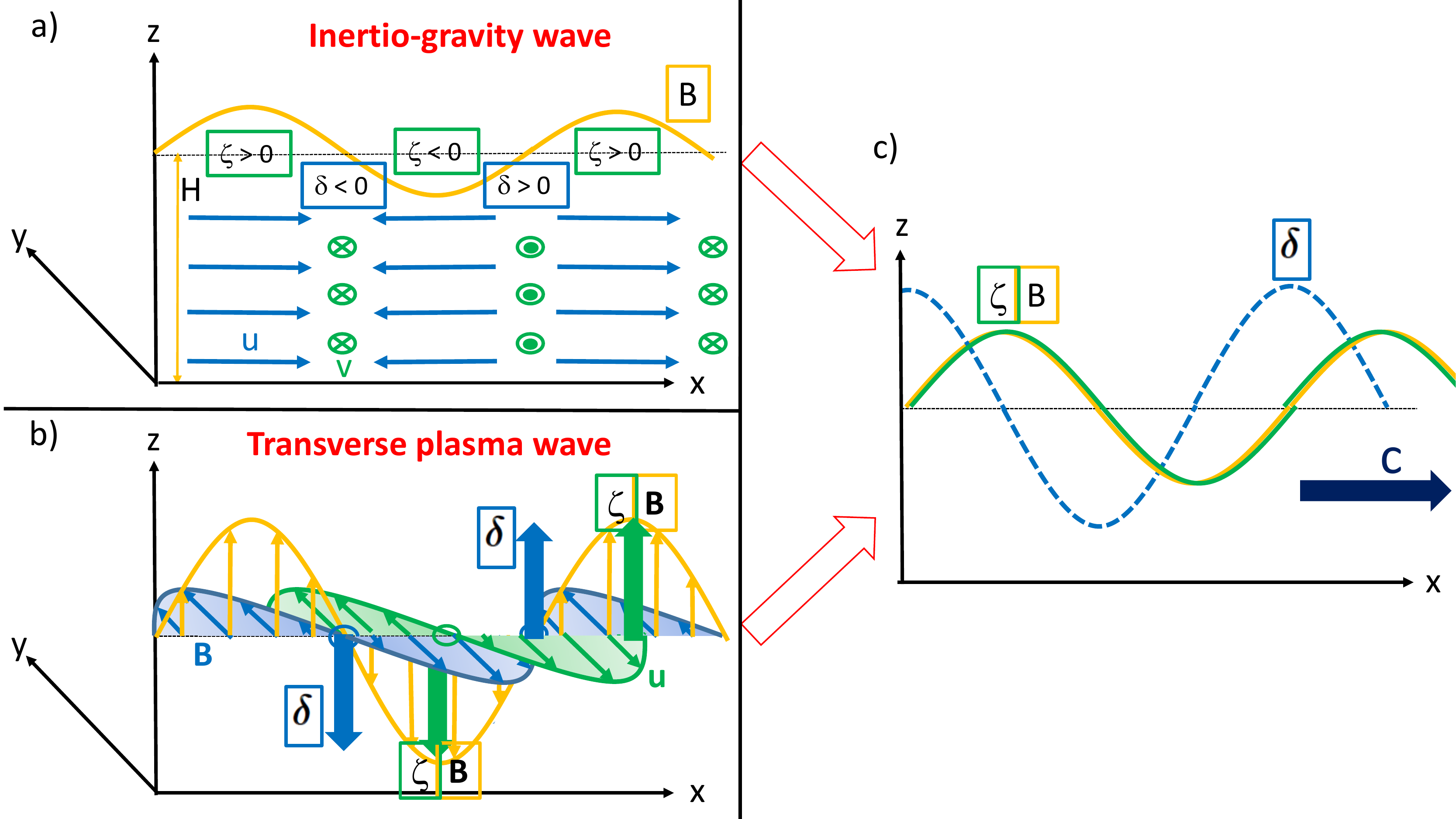}
    \caption{As Fig.\,~\ref{fig:4} for inertio-gravity waves in rotating shallow water and transverse electro-magnetic waves in a cold plasma where electron flow and vorticity fields (green) come into play.
    }
    \label{fig:5}
\end{figure}

Consider now the full systems \eqref{zetaSW}-\eqref{deltaSW} in a rotating fluid  and \eqref{zetaplasma}-\eqref{deltaplasma} in a plasma instead of in vacuum. These admit the inertio-gravity wave and plasma transverse wave solutions respectively, with the equivalent dispersion relations $\omega^2 = f^2 + (kc_s)^2$ and $\omega^2 = \omega_p^2 + (kc)^2$. We first note that, for non-stationary modal solutions, $\zeta = B$ and ${\pmb \zeta} = {\bf B}$. In Fig.\,~\ref{fig:5} we illustrate the wave propagation mechanisms in the two systems. In addition to the gravity wave structure described in Fig.\,~\ref{fig:4}$a$, a rotational velocity field in the $y$ direction generates a vorticity field $\zeta$ that is equal and in phase with the height field $B$. The Coriolis force applied on the rotational velocity part joins the pressure gradient force in translating the divergent velocity part. This explains why the inertio-gravity wave frequency is larger than the gravity one (though the group velocity is smaller than the gravity wave one). By contrast, only the Coriolis force acts on the divergent part to translate in concert the rotational velocity field. Since the Coriolis force is proportional to the velocity field it is applied on, this explains why the divergent field must be larger in magnitude than the rotational part (for leftward propagation the height and the divergent part of the velocity fields should be anti-phased, but the vorticity and the height fields remain in phase).

For transverse electro-magnetic wave propagation in a plasma, the structure described in Fig.\,~\ref{fig:4}b is accompanied by a rotational electron flow field in the horizontal plane, a quarter of a wavelength ahead of the electric field. Consequently, the vorticity field is pointing in the $z$ direction and is in phase with and equal in magnitude to the magnetic field. The new velocity field helps the curl of the magnetic field to translate the electric field, which explains why the frequency of the wave is larger than that of  electro-magnetic waves in vacuum (though the group velocity is obviously smaller than the speed of light in vacuum). By contrast, the electron velocity field is translated only by the electric field (which explains why the curl of the electric field must be larger in magnitude than the vorticity induced by the electron flow). Hence, again, despite  differences in physics, the propagation mechanisms of the two waves, in terms of the wavy structure of the $B$, $\zeta$ and $\delta$ fields, are the same where $|B| =|\zeta|$ and  $\delta$ lags $B$ and $\zeta$ by a quarter of a wavelength.


It is worthwhile to mention that although \eqref{zetaplasma}-\eqref{deltaplasma} is an equation set for vector variables, still in free infinite space and in a Cartesian representation (in which unit vectors are invariant in space and time) the transverse plane waves only admit solutions where the three vectors (${\pmb \zeta}, {\pmb B}, {\pmb \delta}$) are all aligned with the wave phase lines, thus can be practically considered as scalar variables. 
Furthermore, since the plasma system considered here is isotropic (in contrast with the RSW system where both gravity and the system rotation vectors are aligned vertically), the wave phase lines orientation can be identified with the $z$ axis of a Cartesian coordinate system without loss of generality. In this case, both the 3D Laplacian acting on ${\pmb B}$ in \eqref{deltaplasma} and the horizontal Laplacian acting on $B$ in \eqref{deltaSW} become a 1D second-order derivative in the horizontal wave direction of propagation (which is the $x$ direction in Figs.\,~\ref{fig:4}, \ref{fig:5}, for the choice of the electric and the electron velocity fields to be aligned in the $y$ direction). This explains why systems \eqref{zetaSW}-\eqref{deltaSW} and \eqref{zetaplasma}-\eqref{deltaplasma} yield the same plane wave solutions.

It should be also noted that \eqref{zetaSW}-\eqref{deltaSW} and \eqref{zetaplasma}-\eqref{deltaplasma} trivially satisfy the non-trivial solutions: $\zeta = \delta = B = 0$ and ${\pmb \zeta} = {\pmb \delta} = {\bf B} = 0$ of equations sets \eqref{SW_cont_uv}-\eqref{SW_momentum_v} and \eqref{momentumLin}-\eqref{AmpereLin} respectively. These solutions are respectively, the inertial oscillations for the RSW system governed solely by the Coriolis force: 
\be
\der{u}{t} =  fv
\hspace{0.25cm} \& \hspace{0.25cm}
\der{v}{t} =  -fu,
\hspace{0.25cm} \Longrightarrow \hspace{0.25cm}
\omega = \pm f, \hspace{0.25cm} u = - iv\, , 
\label{IW}
\ee
which represents a clockwise circular motion of the velocity vector $(u,v)$ (where both $u$ and $v$ are constant in space),
and the longitudinal plasma oscillations, governed solely by the oscillatory interplay between the irrotational electric and the electron flow fields, lacking a magnetic field: 
\be
\der{{\bf u}}{t} =  -{e\over m} {\bf E}
\hspace{0.25cm} \& \hspace{0.25cm}
\der{{\bf E}}{t} =  {e n_0\over \epsilon_0} {\bf u},
\hspace{0.25cm} \Longrightarrow \hspace{0.25cm}
\omega = \pm \omega_p, \hspace{0.25cm} {\bf E} =  i\omega_p 
\left ({m \over e}{\bf u} \right )\, , 
\label{LP}
\ee
lacking any spatial variation as the oscillations are in phase everywhere.


\section{Some other formal analogies}

We additionally note that, in the fully nonlinear RSW system, the Rossby potential vorticity $q\equiv f(\zeta +1)/h$ (which is not a function of $z$), is materially conserved \citep{vallis}
\be
{D_H\over Dt}q \equiv \left (\der{}{t} + {\bf v}_H\cdot\nabla_H \right) q = 0. 
\label{PVSW}
\ee
Under linearisation, $q = f[1 + (\zeta - B)]/H$, thus \eqref{PVSW}, is reduced to $\partial(\zeta -B)/\partial t = 0$, which implies indeed that for time dependent solutions $\zeta = B$. As pointed out, in the plasma system, it is evident from \eqref{zetaplasma}-\eqref{Bplasma} that the linearised structure of the electro-magnetic waves must satisfy the vector form equality ${\pmb {\zeta}} = {\bf B}$. Hence, it is interesting to examine whether the latter can be obtained from an equivalent version of potential vorticity conversation in the fully nonlinear plasma system. Taking a curl of \eqref{momentum}, and using \eqref{charge_cont}, \eqref{Farady} and \eqref{GaussMagnetic}, we obtain
\be
{D\over Dt}\left [ {(\pmb {\zeta} - {\bf B})\over n}\right ] =\left  [ {(\pmb {\zeta} - {\bf B})\over n}\cdot\nabla\right ]{\bf u}\, ,
\label{PVplasma}
\ee
which is the material line equation for ${(\pmb {\zeta} - {\bf B})/ n}$, where $D/Dt \equiv \left (\partial/\partial t + {\bf u}\cdot\nabla \right)$ is the full 3D material derivative. Hence, only in the restrictive case where the current flow is on a plane, perpendicular to the magnetic field, $(\pmb {\zeta} - {\bf B})/ n$ is materially conserved. Nonetheless, for the linearised dynamics, the quadratic term in the RHS of \eqref{PVplasma} vanishes, and we obtain after linearisation that $n_0^{-1}\partial(\pmb {\zeta} - {\bf B})/\partial t = 0$, which can be regarded as a plasma formal analogue to the linearised potential vorticity conservation law.

There is also a formal analogy between rotating barotropic compressible inviscid flow with shallow water flow \citep{vallis}, and given that the form of \eqref{PVplasma} looks remarkably like the equation for the (Ertel) PV in the compressible system (with $-{\bf B} \leftrightarrow {\bf f}$ since the minus sign results from the fact that upward magnetic field acts to the left of electrons' motion), one might wonder if the linearised plasma system could also be compared to the compressible system. Indeed, the 3D polytropic compressible system is given by \citep{vallis}
\be
\der{\rho}{t} = -\nabla\cdot(\rho {\bf u})\, ,
\label{Pol_cont}
\ee
\be
\der{{\bf u}}{t} = -({\bf u}\cdot\nabla){\bf u}  -{1\over \rho}\nabla p + {\bf u}\times{\bf f}\, 
\label{Pol_momentum}
\ee
with the general polytropic relation $p = a\rho^\gamma$ for the two arbitrary constants $(a,\gamma)$, and ${\bf f} = f{\hat{\bf z}}$, as before. Linearising with respect to a rest state density $\rho_0$, we obtain:
\be
\der{}{t}\left ({\rho'\over \rho_0}\right ) = -
\nabla\cdot{\bf u}'\, ,
\label{Pol_contLin}
\ee
\be
\der{{\bf u}'}{t} = -c_s^2\nabla \left ({\rho'\over \rho_0}\right )  + {\bf u}'\times{\bf f}\, ,
\label{Pol_momentumLin}
\ee
where $c_s^2 = \mathrm{d}p/\mathrm{d}\rho$ is now the square of the speed of sound, considered constant. Denoting the 3D and horizontal divergence by $\delta_3$ and $\delta_H$, respectively, \eqref{Pol_contLin}-\eqref{Pol_momentumLin} can be transformed into:
\be
\der{{\zeta}}{t} = -{{\delta}}_H\, ,
\label{zetaPol}
\ee
\be
\der{{B}}{t} =  -{{\delta}}_3\, ,
\label{BPol}
\ee
\be
\der{{\delta}_3}{t} = -c_s^2\,\nabla_3^2 {B} + f^2\, {{\zeta}}\, .
\label{deltaPol}
\ee
where now $B = \rho'/\rho_0$. While a 3D Laplacian appears now in the RHS of \eqref{deltaPol}, only the horizontal part of the divergence appears in \eqref{zetaPol}. Hence, this system suggests an analogue between electro-magnetic waves in vacuum and sound waves, however the rotating shallow water waves arguably compares better with the cold plasma waves.

The dispersion relation for transverse electro-magnetic waves in plasma has an important consequence in
$\textit{laser}$-produced plasmas. Since $k$ must be a real number for electro-magnetic waves to propagate, a critical value of the plasma frequency $\omega=\omega_c$ determines whether the plasma is transparent ($\omega_l > \omega_c$) or opaque ($\omega_l < \omega_c$) to laser light of frequency $\omega_l$ \citep{chen1974introduction}.
Similarly, due to the increase of $f$ toward the Earth pole, near inertial oceanic inertio-gravity waves (which are ubiquitous waves, generated by wind stress of local storms, whose frequency is close to the low frequency cutoff) tend to propagate equatorward. Super-inertial waves may propagate poleward but then reflect back towards the equator at a nearby turning latitude \citep{Gill}. Hence, in the absence of strong oceanic shear currents and local vortices, the region poleward to that turning latitude remains ``opaque''.


\section{Summary}

Although the cold plasma system (with no background magnetic field) and rotating shallow water systems seem physically disparate, this work highlights a non-trivial formal equivalence between the two systems in the linearised regime. 
The two linearised systems admit a formally equivalent dispersion relation for transverse waves. The vector-scalar analogue suggested here, between the magnetic ${\pmb B}$ and the pressure (height) $B$ fields, and the curl of the electric field ${\pmb \delta}$ and the velocity divergence  field $\delta$, brings the two systems onto  common ground. This allows a similar mechanistic interpretation of the propagation of inertio-gravity waves in the rotating shallow water system and the propagation of electro-magnetic waves in a cold plasma.

The non-trivial analogy is perhaps also interesting in that, if anything, one might have expected the formal analogy to be for the case where there is a vertical background magnetic field in a cold plasma system \citep{Parker-et-al20a}, leading to a non-zero Lorentz force that, at first sight, might be formally identified with the Coriolis force in the rotating shallow water system (with the cyclotron frequency formally identified with the inertial frequency). This is indeed what happens in a warm plasma where the pressure gradient cannot be ignored in the momentum equations and, as a consequence, electron-cyclotron waves  (of frequency below the cyclotron frequency) exhibit features analogous to inertial waves (waves of frequency below the inertial frequency).  However this is not in fact the case in a cold plasma, where we find it is the plasma frequency rather than the cyclotron frequency that  is  formally identified with the inertial frequency. 

It is not the first time formal analogies between plasma dynamics and geophysical fluid dynamics phenomena have been noted. For example, it is well-known that there is a formal equivalence between the Hasegawa--Mima equations and the quasi-geostrophic equations, drift waves and Rossby waves, and the problem of ${\bf E} \times{\bf B}$ drift flow formation and zonal flow formation (see for example \citealt{Diamond-et-al-Plasma} or \citealt{Connaughton-et-al15}, and references therein), which have received notable focus with the view that understanding one system has the potential to advance our understanding in the other. This work provides another example where understanding in geophysical fluid dymamics has implications for plasma dynamics, and it is envisaged there will be further instances where understanding and/or techniques in either research field will have applications in both (e.g. topological waves and topological protection, which is recently gaining traction in geophysical fluid dynamics and plasma physics; see for example \citealt{Delplace-et-al17} and \citealt{Parker-et-al20a}).

As a consequence of the classical dispersion relation of transverse electro-magnetic  waves in plasma, 
laser pulses interacting with a dense, opaque plasma layer ($\omega_l < \omega_c$) initially reflect at its boundary.
However, laser pulses above the \textit{relativistic} intensity threshold (order of $10^{18}\ \mathrm{W}\ \mathrm{cm}^{-2}$, \citep{mourou2006optics}), 
would heat the plasma electrons to nearly the speed of light and thus increase their mass by the Lorentz factor $\gamma = \sqrt{1- v^2 / c^2}$. 
Here the dispersion relation becomes
$\omega^2 = \omega_p^2/\langle\gamma\rangle + (kc)^2$, 
where $\langle\gamma\rangle$  is averaged over the plasma volume. 
This effect, known as relativistic transparency \citep{kaw1970relativistic}, allows intense laser pulses
to interact volumetrically with classically opaque plasmas, to efficiently accelerate electrons and ions \citep{hegelich2013laser}.
An equivalent rotating shallow water system to this case would feature a drop in the effective Coriolis frequency $f_{\rm eff} \equiv f(1 + \overline{\zeta})$,
due to the generation of anticyclonic (negative vorticity) mean shearing current (for instance when $\overline{\zeta} = -\partial \overline u / \partial y <0$, where ${\overline u}(y)$ is a mean, local zonal jet, indicated by an overbar, pointing eastward in the $x$ direction and varies meridionally in the $y$ northward direction). In fact, as suggested by \cite{Tort} this may allow poleward propagation of oceanic near inertial waves beyond turning latitudes defined only by the value of $f$. While highly energetic near inertial waves may alter the mean flow current \citep{Xie}, it would be interesting to find a real-world example by which oceanic near inertial waves alter themselves the mean flow and by that allow further poleward propagation into the otherwise ``opaque region''.\vspace{1cm}

{\bf Acknowledgements}

EH is grateful to Roy Barkan for fruitful discussion. JM acknowledges financial support from the RGC Early Career Scheme 2630020 and the Center for Ocean Research in Hong Kong and Macau, a joint research center between the Qingdao National Laboratory for Marine Science and Technology and Hong Kong University of Science and Technology. 
The order of authorship is alphabetical.
\vspace{1cm}

{\bf Declaration of interest}

The authors report no conflict of interest.


 \newcommand{\noop}[1]{}
  \providecommand{\noopsort}[1]{}\providecommand{\singleletter}[1]{#1}

\end{document}